\documentclass[preprint]{article}

\usepackage[margin=1in]{geometry}

\usepackage{fancyhdr}
\usepackage{amsmath}
\usepackage{amssymb}
\usepackage{easymat}
\usepackage{graphicx}
\usepackage{listings}
\usepackage{cases}
\usepackage{color}
\usepackage{geometry}
\usepackage{pdflscape}
\usepackage[english]{babel}
\usepackage{amsthm}

\usepackage[flushleft]{threeparttable}
\usepackage{bbm}

\usepackage{tikz}
\usetikzlibrary{patterns}
\usepackage{pgfplots}

\usepackage{natbib}
\RequirePackage[colorlinks,citecolor=blue,urlcolor=blue,pdfstartview=FitB]{hyperref}
 
\usepackage{subcaption}
\usepackage{setspace}
\usepackage{multirow}

\title{Inference for Synthetic Control Methods with Multiple Treated Units}
\author{Ziyan Zhang\footnote{Cambridge University, zz365@cam.ac.uk.}}

\begin{document}

\maketitle

\abstract{Although the Synthetic Control Method (SCM) is now widely applied, its most commonly-used inference method, placebo test, is often problematic, especially when the treatment is not uniquely assigned. This paper discuss the problems with the placebo test under multivariate treatment case. And, to improve the power of inferences, I further propose an Andrews-type procedure as it potentially solve some drawbacks of placebo test. Simulations are conducted to show the Andrews' test is often valid and powerful, compared with the placebo test.}

\section{Introduction}

When estimating the treatment effect of multiple treatment units using the synthetic control method, the current literature often considers the placebo test in testing the significance of the average treatment effect. However, the validity of this test is questionable in this case, mostly because the relatively large number of the treatment units would lead to imprecise null distribution thus undermine the power of the test. 

\cite{crest2018penalized} discussed the modification of synthetic control estimation method for multiple treatment case. They choose the weight for control units by considering both the pre-treatment fit and the discrepancy of controls to reduce the potential bias caused by interpolation. When it comes to the inference, they apply the permutation test to the ratio of aggregate mean square error between pre-and post-treatment periods. However, there are two main issues that may disturb the performance of the test. The heterogeneity in sample units might lead to over-rejecting, and the high probability of including treated units tends to impair the power of the test.

Thus, to improve the potential issues with the placebo test, I apply Andrews' test which is introduced to assess the instability at the end of a single series in \cite{Andrews2003}. The test uses prediction errors to constitute statistics, which is the average discrepancy between treated group and “synthetic” control group in our case. The instability will present if the average treatment effect exists.

Although I only consider the case when there is only one post-treatment period, the inference method can be further extended to cases with multiple post-treatment periods. The single post-treatment case is a simplified version of multiple periods thus the issue could be solved by repeatedly implement the procedures. 

This paper is related to a literature that receives increasing attention. Ever since \cite{Abadie2003} first introduced the synthetic control method (SCM) and applied it to estimate the treatment effect of terrorist activities, there are a large number of papers later followed the manner and attempted to improve the method thus overcoming the limitations of this method. \cite{Doudchenko2016} and \cite{Li2019} applied modified SCM by adding an intercept and abandon the restriction that all weights sum to one. The method help to find solutions to the case when the controlled units cannot perfectly form a group that is at the same level of the treated group. Similarly, \cite{Ferman2016}, in order to keep the treated group and the control group at a relatively same height, they did the estimation after demeaning the counterfactual by the actual data. It was showed that the estimator can improve the bias and variances compare to the traditional estimators. On the other hand, \cite{crest2018penalized} added a penalty parameter to the original SCM estimator and showed it outperforms estimators using other estimation methods for the multiply treatment case. For the overview of the SCM, \cite{Doudchenko2016} concluded main modifications and distinguished proper methods for various situations.

Just as \cite{athey2016state} acclaimed that SCM is the most important innovation regarding the policy evaluation method in the past 15 years, the SCM is now widely applied to the evaluation of treatment in many areas. The point that is inadequately discusses in the wide range of those applications, however, is the issue with inference. Although \cite{Abadie2003}, \cite{Abadie2010} and \cite{Abadie2015} papers introduced the permutation test and kept on improve the test by using different test statistics and validation method, the amount of literature that question the credibility of the test and attempted to improve the inference is large. 

I mainly focus on the inference of synthetic control methods. \cite{Hahn2017} simulated several settings of the permutation test and raised the concern that the placebo test would likely to fail if there are aggregate shocks present. Although they discussed some potential solutions but not solve the inference issue indeed.  Additionally, regarding the issue of inference of multivariate case, \cite{Firpo2018} tried to modify the inference procedures by assigning different probabilities to the treated unit using explanatory variables when computing the $p$-value as well as alter the formation of test statistics. \cite{cao2019estimation}, on the other hand, turned to study the variation between different periods and tested the treatment effect and spillover effect using Andrews’ test which first introduced in \cite{Andrews2003} paper and further discussed in \cite{Andrews2006}.  More papers such as \cite{Chernozhukov2017} combined the inference of synthetic control method with many other policy intervention evaluation methods and provided feasible testing methodologies. Based on all those past papers, we would like to study how the testing approaches perform when the treated unit is not unique. For this paper, we focus on the inference of the treatment effects. For inference problems regarding the synthetic control weights, one may refer to the literature on the inference problem with restricted parameter space. For example, See \cite{ANDREWS1998155}, \cite{doi:10.1111/1468-0262.00210}, and \cite{yu2019constrained}.

The rest of this paper will be arranged as following. 
Section \ref{model_section} illustrates the estimation method. In section \ref{inferece_section} I propose the inference method that are used frequently past and the modification of recently introduced method. Section \ref{sim_section} verifies our advocating of Andrews' test. Section \ref{conclusion_section} concludes.

\section{Model and estimation }
\label{model_section}

\subsection{Setup}
With denoting \textit{D} the dummy of treatment and $D=1$ when the unit is treated, we following Rubin’s (1974) notation of the potential outcomes of the treated units $Y_{i,t}$ as 
\begin{equation}
	Y_{i,t}=\begin{cases}
		Y_{i,t}(1), &\text{if} \quad D=1\\
		Y_{i,t}(0), &\text{if} \quad D=0\\
	\end{cases}
\end{equation}
Define the potential outcome of non-treated units as $Y_{j,t}$, the expression of it is the same as for $Y_{i,t}$.
Suppose that $i$ belongs to the first $n_1$ units out of the total $N$ units and $j$ belongs to the following $n_0$ non-treated units, there is no treatment in first $T_0$ periods thus define it as the pre-treatment period. For simplicity, assume we only observe 1 post-treatment period. Extension to multiple post-treatment periods case is straightforward by analysing each period individually following the steps of the single case. The treated units will show treatment effect $\alpha_{i}$ only during the post-treatment period,  By the nature of treatment, however, we could only observe the actual outcome of treated units under treatment at the post period but cannot directly observe how it would be like if no treatment. We thus estimate the treatment effect by the outcome of the synthetic control group following the method introduced in \cite{crest2018penalized} in later section. In this paper, we consider the case of moderate sample size, for which the number of units $N$ considered is approximately equal to or smaller than the number of time periods $T$.

\subsection{Estimation}
There are plenty of synthetic control estimation methods following \cite{Abadie2003} paper which first introduced this estimation method. For example, the modified synthetic control method by \cite{Li2019} and the demeaned SCM method by \cite{Ferman2016}. Since we concentrate on the inference method, the computation method of the synthetic control weights is not essential to our analysis. Here we apply the method introduced in \cite{crest2018penalized}, since it is suited to the case of multiple treatments. 

According to \cite{crest2018penalized}, the method of penalized synthetic control gives better estimation since it assigns penalty to controls that are distant. Namely,for each $i$, we solve the following optimization problem:
\begin{equation}
	\label{min}
	\smash{\displaystyle\min_{W{ij}\in R}}\quad \sum_{t=1}^{T_{0}} \left(Y_{i,t}-\sum_{j=n_{1}+1}^{N} W_{ij}Y_{j,t} \right)^{2} + \gamma \sum_{t=1}^{T_{0}}\sum_{j=n_{1}+1}^{N} W_{ij}\left(Y_{i,t}-Y_{j,t} \right)^{2}
\end{equation}
\begin{equation*}
	\text{subject to}\quad W_{ij}\geq 0, \quad \sum_{j=n_{1}+1}^{N}W_{ij}=1,
\end{equation*}
where $W_{ij}$ are the weights of the $j$th control to the $i$th treated unit. $Y_i$ and $Y_j$ are the vectors of treated and control units respectively. Solving \eqref{min} gives the synthetic control weights, we then sub them in to the post-treatment period thus calculate the average treatment effect across all treated units. As the weight would be affected by the penalty parameter $\gamma$, the synthetic control estimator is:
\begin{equation}
	\label{error}
	\widehat{u}_{i,T_{0}+1}=\frac{1}{n_{1}}\sum_{i=1}^{n_{1}}\left[Y_{i,T_{0}+1}-\sum_{j=n_{1}+1}^{N}W_{ij}^{*}(\gamma)Y_{j,T_{0}+1} \right]
\end{equation}
Here $W_{ij}^*(\gamma)$ is the optimal weight after iterations and finding the penalty parameter which minimises the mean square error. 
The average estimation error of the pre-treatment period can be computed in the same way. 
Additionally, the penalty parameter here is selected by dividing pre-treatment data into training part and validation part. For each value of $\gamma$, we calculate the weights of controls in the training periods and calculating the mean squared error (MSE) applying the weights to the validation period. The $\gamma$ gives minimum MSE will be used then to compute the weights when include all pre-treatment data.

\section{Inference}
\label{inferece_section}

The core consideration of this paper is to compare the sizes and powers of different tests under various settings of data. We concern about it due to the speciality of multiple treatment case. Although currently the most widely applied testing methodology is placebo test as developed in \cite{Abadie2003}, the method requires homogeneity of units and is originally developed for single treatment situation. For the multiple treated units situation, such as randomly select a group of people or cities to examines if some specific policies are applicable, it is highly unlikely the treated units will be selected simply by randomization, thus we are interested in finding out if there are alternative test techniques that do not require such strong assumptions. 

Here in this paper, we consider the test used in \cite{crest2018penalized}, which tests the aggregate effect exhibited by all treated units. While the placebo test check the variation across units, another test method as practiced in \cite{cao2019estimation} concentrated on the variation of time within the dataset. In the so called Andrews’ test, we test if the post-treatment period show different patterns compare to the pre-treatment periods.  We are going to study the pros and cons of both testing methodologies and suggest which is more suitable for different cases.

\subsection{Current methods}

Although the mechanism of placebo test is fixed, the statistic used to implement the test is different. In this paper,we refer from \cite{crest2018penalized}, and use the ratio of mean squared prediction error (RMSPE) to test if the ratio of the post-treatment period is abnormal with respect to the null distribution. The statistic is expressed as the ratio of the pre- and post-treatement data,
\begin{equation}
	\widehat{T}=\frac{\sum_{t=1}^{T_{0}}\left(\sum_{i=1}^{n_{1}}\widehat{u}_{i,t} \right)^{2}}{\left(\sum_{i=1}^{n_{1}}\widehat{u}_{i,T_{0}+1} \right)^{2}}
\end{equation}
Here the $\widehat{u}_{i,t}$ is the prediction error of each treated period for each treated unit, and it is equal to $Y_{i,t}-\sum_{j=n_{1}+1}^{N}W_{ij}^{*}(\gamma)Y_{j,t}$. Theoretically, if the treatment effect exist, the denominator, which represent the post-treatment aggregate prediction error, will be larger than the case of without treatment, thus the whole ratio should be smaller. This permutation-type test is then to simply randomly select $n_1$ units out of all treated and untreated units, and the $n_1$ units that selected would then be regarded as the assuming ‘treated’ units. For every randomization of ‘treated’ units, we compute the statistic $T(p)$ after $p$th time redefining the treatment group, and aggregating all the $T(p)$ values will form the null distribution for the statistic under the sharp hypothesis $H_0: \alpha_{1}=\alpha_{2}=...=\alpha_{n_{1}}$. Suppose we repeat the step of randomly select ‘treated’ units and calculate $T(p)$ for $P$ times, with the calculated test statistic and the null distribution, the $p$-value $p_{placebo}$ is then           
\begin{equation}
	\frac{1}{P+1}\left\lbrace 1+\sum_{p=1}^{P}\mathbb{1} \left[ T(p)\leq \widehat{T} \right] \right\rbrace
\end{equation}
We reject the null $H_0$ when $p_{placebo}\leq\alpha$ with $\alpha$ being the significance level. 

The placebo test sets a baseline of the test result, but the result are questionable mainly out of the concern of two issues. The first question is that if there exist heterogeneity across units, placebo test tend to over-rejecting the null. While, as discussed in \cite{cao2019estimation}, Andrews' test does not require homogeneity across units but only care about the homogeneity across time. 

Another issue concerns about the speciality of multiple treatment case. Since the placebo test establish its null distribution by repeatedly and randomly select $n_{1}$ 'treated' units for many times, under the multivariate case, the probability that selecting a truly treated units in the repeating rounds is rising fast, thus the null distribution will move away from the true distribution and therefore reduce the power of the test. The issue presents since the placebo test makes use of the variation between units while the limited number of the treated units will make distinguishing abnormal performance easily. When the treated number increases, it makes recognizing the treated effect harder. We refer to Andrews' test to solve this problem since it cares variations between time periods, and treatment effect only presents in the specific period after treatment thus multiple treated units would not affect the test result. As Andrews' test simultaneous solve both issues, we expect it will have better performance. We concentrate our study in the second issues in this paper, since many past papers have discussed the above question.

\subsection{Andrews' test}
In this test, we are going to use the prediction errors $\widehat{u}_{i,t}$ and $\widehat{u}_{i,T_{0}+1}$ as mentioned in previous sections. By assumption, the prediction error $u_t$ is stationary with mean zero, we can thus test if there are instabilities at the end of the process, i.e., instability at the post-treatment period.

The most straightforward method is to test if the sum of prediction error of the post-treatment period is at the tail of the null distribution. We could then define the test statistic $\widehat{S}$ to be
\begin{equation}
	\sum_{i=1}^{n_1}\left(\widehat{u}_{i,T_{0}+1}  \right)^2
\end{equation}
It tests if there are treatment effect in one of the post-treatment periods that we interest in. Similarly, the null in this case is also $H_0: \alpha_{1}=\alpha_{2}=...=\alpha_{n_{1}}$ and the null distribution constituted by all $S(t)=\sum_{i=1}^{n_1}\left(\widehat{u}_{i,t} \right)^2,\text{for each}\quad t=1,2,..., T_0 $, using which we could calculate the $p$-value of this simple Andrews' test $p_{andrew}$ which equals
\begin{equation*}
	\frac{1}{T_{0}} \sum_{t=1}^{T_0}\mathbb{1}\left[S(t)\geq \widehat{S} \right]
\end{equation*}
In other words, $p_{andrew}$ is the average value of the indication function $\mathbb{1}\left[S(t)\geq \widehat{S} \right]$.

Consider the mechanism of the simple Andrews' test, one important issue is the correlation presents in the estimation. Usually, we do not hope to apply the estimations to the data that used in the estimation itself, as the estimated results are surely correlated to the data in this case. Since the correlation might complicates the inference thus gives an incorrect rejection rate, we improve the Andrews’ test in part $(i)$ by using the leave-one-out method for estimation and thus correct the errors as much as possible \citep{Andrews2003}. The leave-one-out estimation helps to correct the finite sample bias which improves the estimation of errors. Then, using the corrected errors, we calculate the $p$-value of the test.

For the leave-one-out estimation, we solve the question in (1) using all pre-treatment data except for the period $t$, where $t=1,2,…,T_{0}$. With the estimated weights, compute $\widehat{u}_{i,-t}=Y_{i,-t}-\sum_{j=n_{1}+1}^{N}W_{ij}^{-t}(\gamma)Y_{j,-t}$. Here $Y_{i,-t} =\left[Y_{i,1} ,Y_{i,2} ,…,Y_{i,t-1 } ,Y_{i,t+1 } ,…Y_{i,T_0} \right]$, similar for $Y_{j,-t}$. Using the corrected error $u_{i,-t}$, we then compute the test statistics and calculate the $p$-value $p_{al}$, which is different from the one simply using the weights from the estimation section.

Let $\tau$ be the significance level and $\phi=1$ if the hypothesis is rejected and 0 otherwise. The validity of the test is established as 
\newtheorem{theorem}{Theorem}
\begin{theorem}
	\label{Theorem}
	Under the sharp null hypothesis that all treatment effects are $0$, $E[\phi]\rightarrow \tau$ as $N\rightarrow \infty$
\end{theorem}
The usage of this Andrews-type testing procedure is theoretically justified in \cite{cao2019estimation} and skipped here.


\section{Simulation}
\label{sim_section}

\subsection{Data Generating Process (DGP)}
The DGP of the simulation follows the setup of \cite{cao2019estimation} paper. As we consider a factor model of the expected outcomes $y_{i,t} (0)$ and $y_{j,t} (0)$, for treated units, the expression of the outcomes is:
\begin{equation}
	\label{factor_model}
	y_{m,t} (0)=\eta_t+\lambda_t^{'} \mu_m+\varepsilon_{m,t}
\end{equation}
where $m$ representing all $i$ and $j$. Thus Equation \eqref{factor_model} indicates that the expected outcome for all the units are generated following same manner, but the observed outcome are computed differently. For non-treated units, $y_{j,T_{0}+1}=y_{j,T_{0}+1}(0)+e_{j,T_{0}+1}$ while for the treated units $y_{i,T_{0}+1}=y_{i,T_{0}+1}(0)+\alpha_{i}+e_{i,T_{0}+1}$. When there is no treatment effect, i.e., $\alpha_{i}=0$, the observed value of treated units should be similar to the untreated units.

Along with the baseline model, we further generate the elements needed in the model, that is, for $ \lambda_{t} =\left(\lambda_{1,t} ,\lambda_{2,t}  ,\lambda_{3,t}\right)'$
\begin{align*}
	&\eta_{t}= 1+0.5\delta_{t-1}+\nu_{0,t} \\
	&\lambda_{1,t} = 0.5\lambda_{1,t-1}+ \nu_{1,t} \\
	&\lambda_{2,t} = 1+\nu_{2,t}+0.5\nu_{2,t-1} \\
	&\lambda_{3,t} = 0.5\lambda_{3,t-1}+\nu_{3,t}+0.5\nu_{3,t-1}
\end{align*}
The error terms, $\varepsilon_{m,t}$ and $\nu$ are i.i.d. standard normal. The individual effect term $\mu_{m}$ is drawn from the $U[0,1]$ distribution and is fixed for each loop. This is the case where all elements are stationary thus Andrews’ test is applicable. For the error process, we first generate it assuming homogeneous errors, since homogeneity is necessary for placebo test to be available. It could later be transformed to be heterogeneous thus to study if the transformed Andrews’ test will outperform placebo test under both cases of homogeneity and heterogeneity. 

After generating the potential outcomes for all units, the observed outcomes will be accordingly generated. We set the treatment effect $\alpha$ to be from 0 to 3 and calculate the rejection rate of each test for different treatment effect cases. Note that the penalty parameter $\gamma$ to be 0.2 for simplicity, the value is estimated before the loop and although for different dataset the parameter will change a bit, it is around 0.2. 
\subsection{Results}
\begin{figure}[h!]
	\centering
	\begin{subfigure}{1\textwidth}
		\centering
		\caption{$(T_{0},n_1,n_0)=(50,10,20)$}
		\includegraphics[scale=.5]{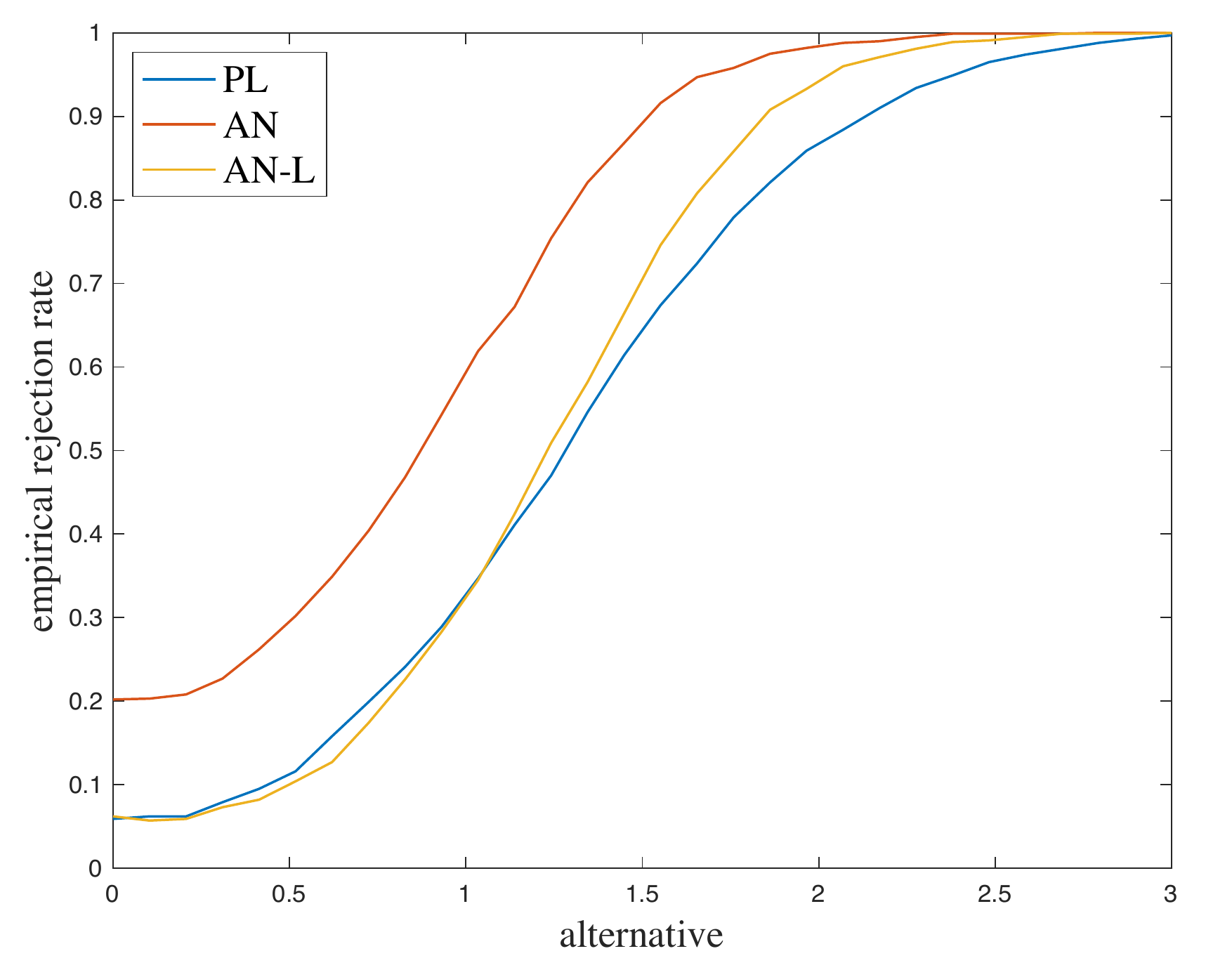} 
		\label{fig:subim1}
	\end{subfigure}
	\begin{subfigure}{1\textwidth}
		\centering
		\caption{{ $(T_{0},n_1,n_0)=(50,30,30)$} }
		\includegraphics[scale=.5]{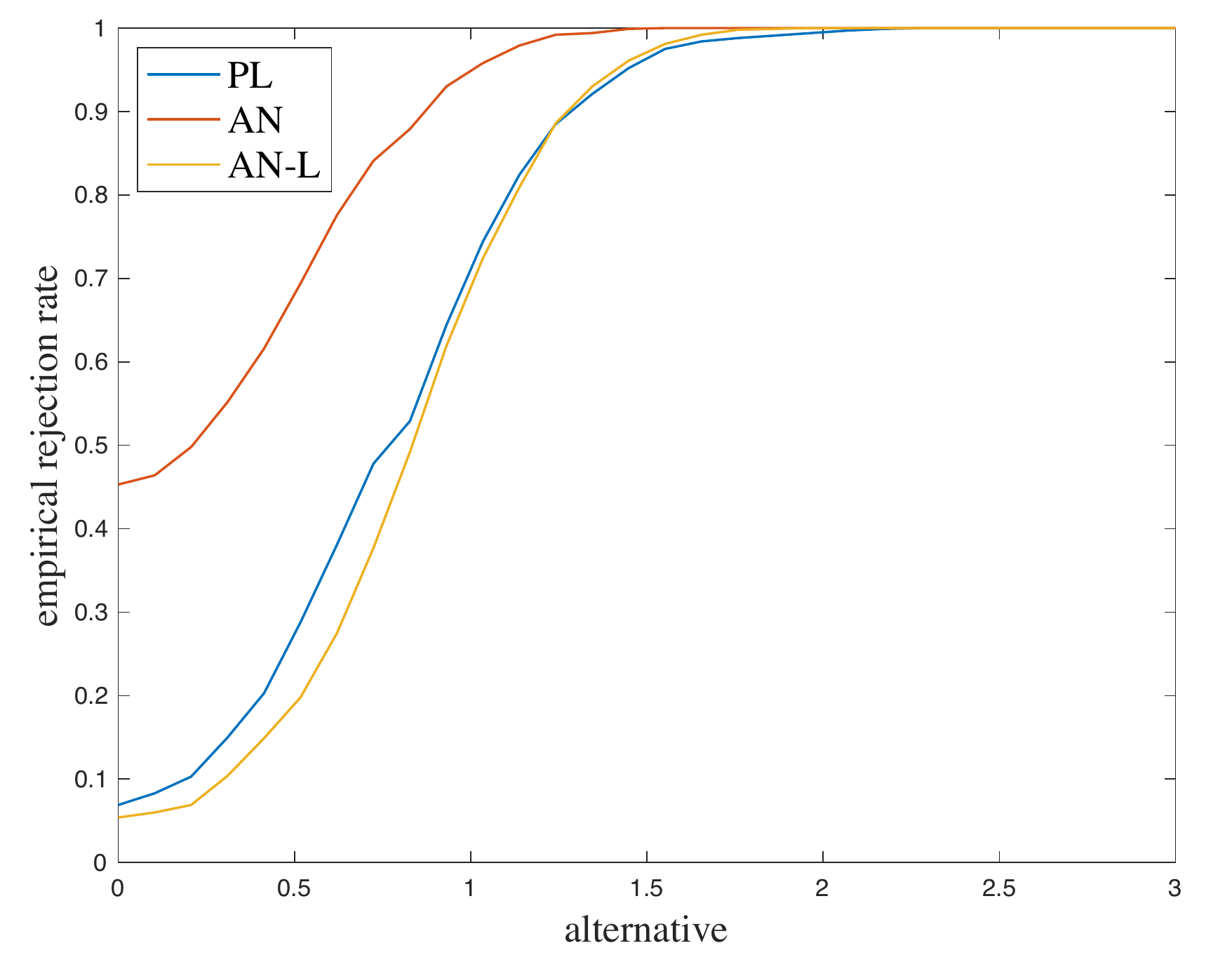} 
		\label{fig:subim2}
	\end{subfigure}
	\caption{Empirical Rejection Rates with different alternatives}
	\label{fig:image2}
\end{figure}
The graphs of simulation results of two data settings are shown in Figure \ref{fig:subim1} and \ref{fig:subim2}. \\

In general, placebo test is robust in terms of sizes and Andrews' test (with leave-one-out procedure) is more powerful.
Specifically, the leave-one-out method clearly helped control the size of Andrews’ test. The rejection rate of Andrews’ test with leave-one-out (AN-L) estimation is around 0.05 for all settings thus it is a valid test. 

Figure 1 show the rejection rates of two tests with different settings. In \ref{fig:subim1}, it is clear to see that AN-L is correctly sized while has higher power than placebo test. While in \ref{fig:subim2} as both tests perform well in size, the AN-L has no obvious difference compare to placebo test. This is due to in setting of \ref{fig:subim2}, the amount of sample units outnumbers time periods $T$, as Andrews' test uses variation in time to construct the null distribution, in this case the number of $T$ may not enough to recover the true null distribution. Apart from placebo test and AN-L test, we also included ordinary Andrews' test. In both cases it over-reject the null due to the correlations between estimation and data.

\section{Conclusion}
\label{conclusion_section}

In this paper, we studied the sizes and powers of two different inference methods and found that in the case of moderate sample size, if there are more than one treated units, the placebo test might be less powerful than Andrews' test (with leave-one-out procedure). The issue presents due to the high possibility that a treated unit is more likely to be included into the calculation of null distribution thus makes it shift away from the true distribution. Andrews' test help to improve the power of the test when there are sufficient number of pre-treatment time periods. The comparison of the two tests indicate that both of the tests have their more suitable application cases. Here in this paper, we verified that Andrews' test is valid and has higher power with multiple treatment case when the number of time periods $T$ is sufficiently large compare to the number of sample units $N$. 


\bibliographystyle{apalike}
\bibliography{ref}

\end{document}